\documentclass[aps,pra,twocolumn,groupedaddress]{revtex4}
\usepackage{graphicx}

\usepackage{ulem}

\begin{document}

\title{Passive harmonic mode-locking by mode selection in Fabry-Perot diode lasers with 
patterned effective index}

\author{D. Bitauld, S. Osborne, and S. O'Brien}

\affiliation{Tyndall National Institute, University College, Lee Maltings, Cork, Ireland}

\begin{abstract}

We demonstrate passive harmonic mode-locking of a quantum well laser diode designed 
to support a discrete comb of Fabry-Perot modes. Spectral filtering of the mode 
spectrum was achieved using a non-periodic patterning of the cavity effective index.
By selecting six modes spaced at twice the fundamental mode spacing, near-transform 
limited pulsed output with 2 ps pulse duration was obtained at a repetition rate of 
100 GHz.

\end{abstract}

\maketitle

The generation of high power, ultrashort optical pulses using mode-locked 
diode lasers is of considerable importance across a range of applications in 
optical communication systems and data processing \cite{avrutin_00, williams_04}. 
In recent years, the wide gain bandwidth of certain self-assembled gain 
materials has enabled sub-picosecond and femtosecond pulsewidths to be obtained 
\cite{rafailov_05, lu_08, merghem_09}. In parallel, various techniques such as 
colliding pulse mode-locking (CPML) and harmonic mode-locking have been adapted for 
diode lasers that allow higher repetition rates to be achieved while avoiding the 
lower average power associated with short device lengths \cite{avrutin_00}. 

CPML places the saturable absorber section at the center of the cavity. 
Two counterpropagating pulses are formed and collide at the center. In this way 
the saturation of the absorption is enhanced and the repetition rate is simultaneously 
doubled from the fundamental \cite{chen_91}. Harmonic mode-locking of diode lasers can 
be achieved by forcing the device to lock on a set of equally spaced but 
non-adjacent cavity modes. With this approach, terahertz (THz) repetition rates 
have been achieved in devices that incorporate distributed Bragg reflector 
mirrors \cite{arahira_94} and intra-cavity reflectors \cite{avrutin_02}. A 
limitation of these approaches to harmonic mode-locking however is the fact 
that only the harmonic frequency rather than the precise form of the mode-locked 
spectrum itself can be specified.  

In this letter we demonstrate an approach to passive mode-locking of diode lasers 
based on the selection of individual modes from the spectrum of Fabry-Perot resonances. 
Mode selection is achieved by a distributed reflection mechanism that is designed as 
a perturbation of the Fabry-Perot mode spectrum. Such an approach has some features 
in common with that described in \cite{avrutin_02}, although, in our case the 
effective index profile along the device is derived directly from an inverse 
problem solution. This approach has allowed us to design single-mode \cite{obrien_05}, 
and two-mode Fabry-Perot diode lasers \cite{obrien_06} with high spectral purity and 
can in principle provide a much greater degree of control of the mode-locked spectrum 
of the laser. To demonstrate passive mode-locking of a discrete comb 
of Fabry-Perot modes, a saturable absorber section is placed adjacent to one of 
the cavity mirrors. In the device we consider here, six primary modes are chosen 
with a spacing of two fundamental modes leading to mode-locking at the first 
harmonic of the cavity. 

We have shown that a set of self-consistent equations for the lasing 
modes can be found by making an expansion about the cavity resonance condition in a 
Fabry-Perot laser \cite{obrien_06b}. The effective index along the Fabry-Perot cavity 
is modified by $N$ additional features that are described by an index step 
$\Delta n$. In our case, these features are slotted regions etched into the ridge 
waveguide of the device. We assume that the index step, $\Delta n$, associated 
with the etched features is a real quantity, although some scattering losses are
inevitably introduced that strongly depend on the depth of the slotted region 
\cite{lu_06}. The threshold gain, $\gamma_{m}$, can 
then be expressed at first order in the index-step as $\gamma_{m} = \gamma_{m}^{(0)} 
+ (\Delta n/n) \gamma_{m}^{(1)}$. Here $\gamma_{m}^{(0)}$ are the mirror 
losses of the unperturbed cavity and the function $\gamma_{m}^{(1)}$ 
describes the effect of the coupling of each feature to the cavity mirrors.
We choose a particular mode $m_{0}$ as an origin in wavenumber space. If we then 
assume that the features are positioned such that at the frequency of mode $m_{0}$, 
a half wavelength subcavity is formed between each feature and one of the cavity mirrors, 
this function takes a particularly simple form:
\begin{eqnarray}\label{gamma_m1}
&\gamma_{m}^{(1)} = \frac{1}{L_{c}\sqrt{r_{1}r_{2}}}\cos(m_{0}\pi) \cos(\Delta m\pi)  \nonumber \\
& \times \sum_{j=1}^{N}A(\epsilon_{j}) \sin(2\pi\epsilon_{j}m_{0}) \cos(2\pi\epsilon_{j}\Delta m).
\end{eqnarray}
In the above expression, the factor $A(\epsilon_{j}) = r_{1} \exp(\epsilon_{j}L_{c}\gamma_{m}^{(0)}) 
- r_{2} \exp(-\epsilon_{j}L_{c}\gamma_{m}^{(0)})$ and $\epsilon_{j}$ is the fractional position of 
the center of each feature measured from the center of the cavity. 

\begin{figure}[t]
\centerline{\includegraphics[width=8.0cm]{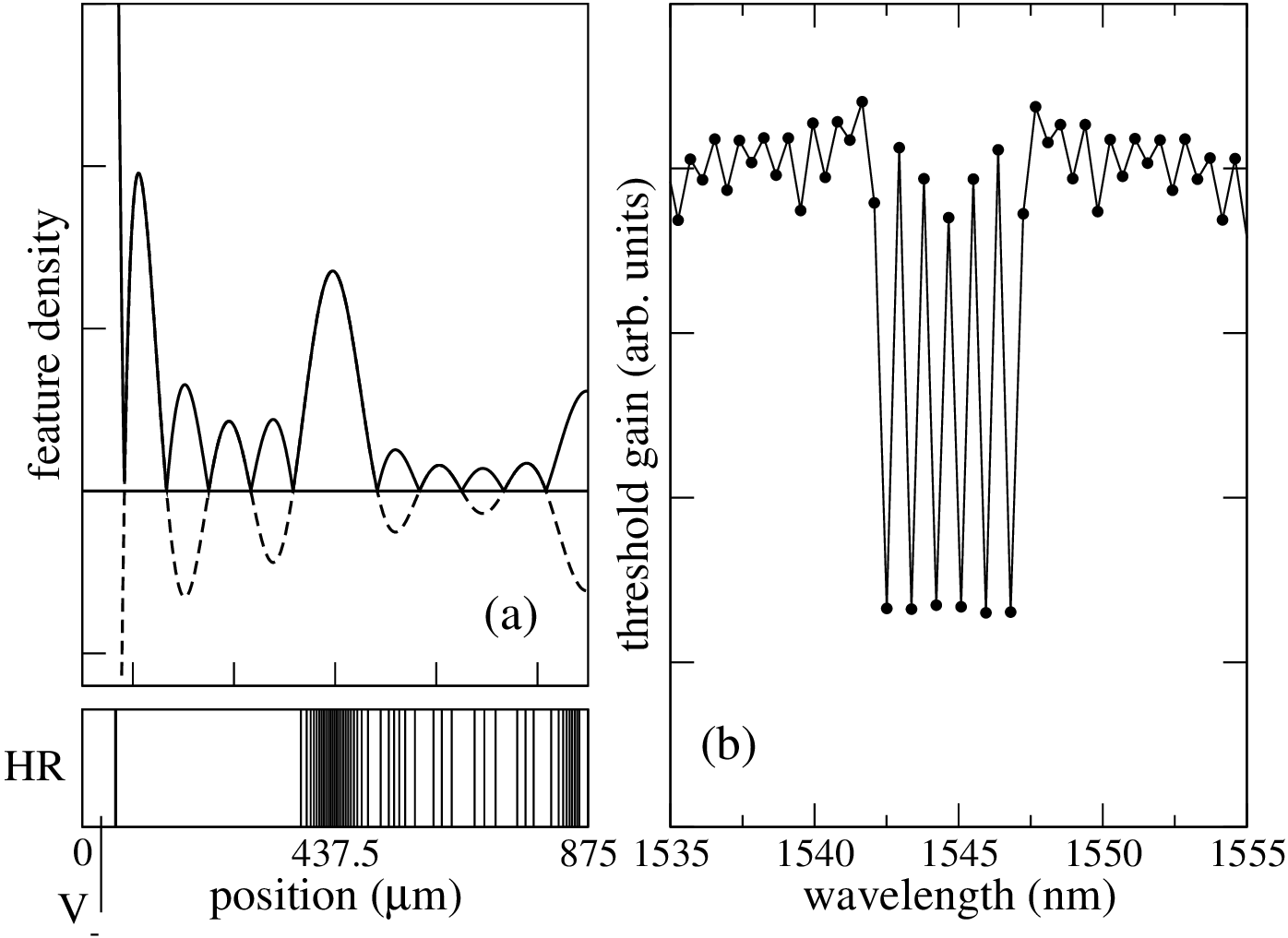}}
\caption{\label{ob_ft} (a) Feature density function (solid line). Dashed lines 
indicate intervals where the Fourier transform of the spectral filtering function 
chosen is negative. (b) Calculation of the threshold gain of 
modes for the laser cavity schematically pictured in the lower panel of the figure. 
Lower panel: Laser cavity schematic indicating the locations of the additional features. 
The device is high-reflection (HR) coated and includes a saturable absorber section adjacent
to the HR mirror as indicated.}
\end{figure}

Eqn. \ref{gamma_m1} above allows Fourier analysis to be used 
to make a direct connection between the index profile in real space and the 
threshold gain modulation in wavenumber space. Here we consider a 
device with six primary modes and a primary mode spacing of $a = 2$ fundamental cavity 
modes (first harmonic). Neglecting frequency dispersion, the appropriate filtering function 
is then a series of equally spaced sinc functions, as each sinc function naturally 
selects a single mode while leaving all others unperturbed. Selecting modes in pairs 
that are centered at the origin chosen in wavenumber space, the Fourier transform of the 
threshold gain modulation function is $\cos(\pi a \epsilon) + \cos(3\pi a \epsilon) + 
\cos(5\pi a \epsilon)$ for $-1/2 \leq \epsilon \leq 1/2$ and is zero otherwise. 
Multiplying this Fourier transform function with the envelope function $[A(\epsilon)]^{-1}$,
we obtain the feature density function shown in Fig. \ref{ob_ft} (a). 

The feature density function shown is then sampled in order to define the appropriate 
effective index profile \cite{obrien_06b}. The cavity is asymmetric with one high-reflection
coated mirror, and features are placed on the opposite side of the device center to this 
mirror where the feature density function is more uniform. This is possible because only 
positive Fourier components are required to form the desired threshold gain variation in the
case considered. Note also that at each zero of the 
feature density function, a $\pi/2$ phase shift is introduced into the effective 
index profile. The low-density grating structure that results is then related to a 
superstructure Bragg grating that is modulated by a series of lower spatial frequencies. 
A schematic picture of the device, high-reflection coated as indicated, is shown in 
the lower panel of Fig. \ref{ob_ft} (a), where $N = 48$ 
etched features are introduced. The calculated form of the threshold gain spectrum 
is shown in Fig. \ref{ob_ft} (b). One can see that six modes are selected around 
a central wavelength of 1545 nm. The comb of selected modes spans a bandwidth in 
frequency of 500 GHz for a device length of 875 $\mu$m.

We now present experimental data obtained from a ridge waveguide Fabry-Perot laser 
fabricated to the design depicted in Fig. \ref{ob_ft}. These data serve to illustrate 
how the mode selection mechanism plays an important role in two distinct 
dynamical states of the device, which is a multi-quantum well InP/InGaAlAs laser 
with a peak emission near 1.5 $\mu$m. A saturable absorber section of length 60 
$\mu$m was placed adjacent to the high reflectivity mirror. During the measurements 
the temperature of the device was stabilized at $15^{\circ}\mbox{C}$.  

\begin{figure}[t]
\centerline{\includegraphics[width=8.0cm]{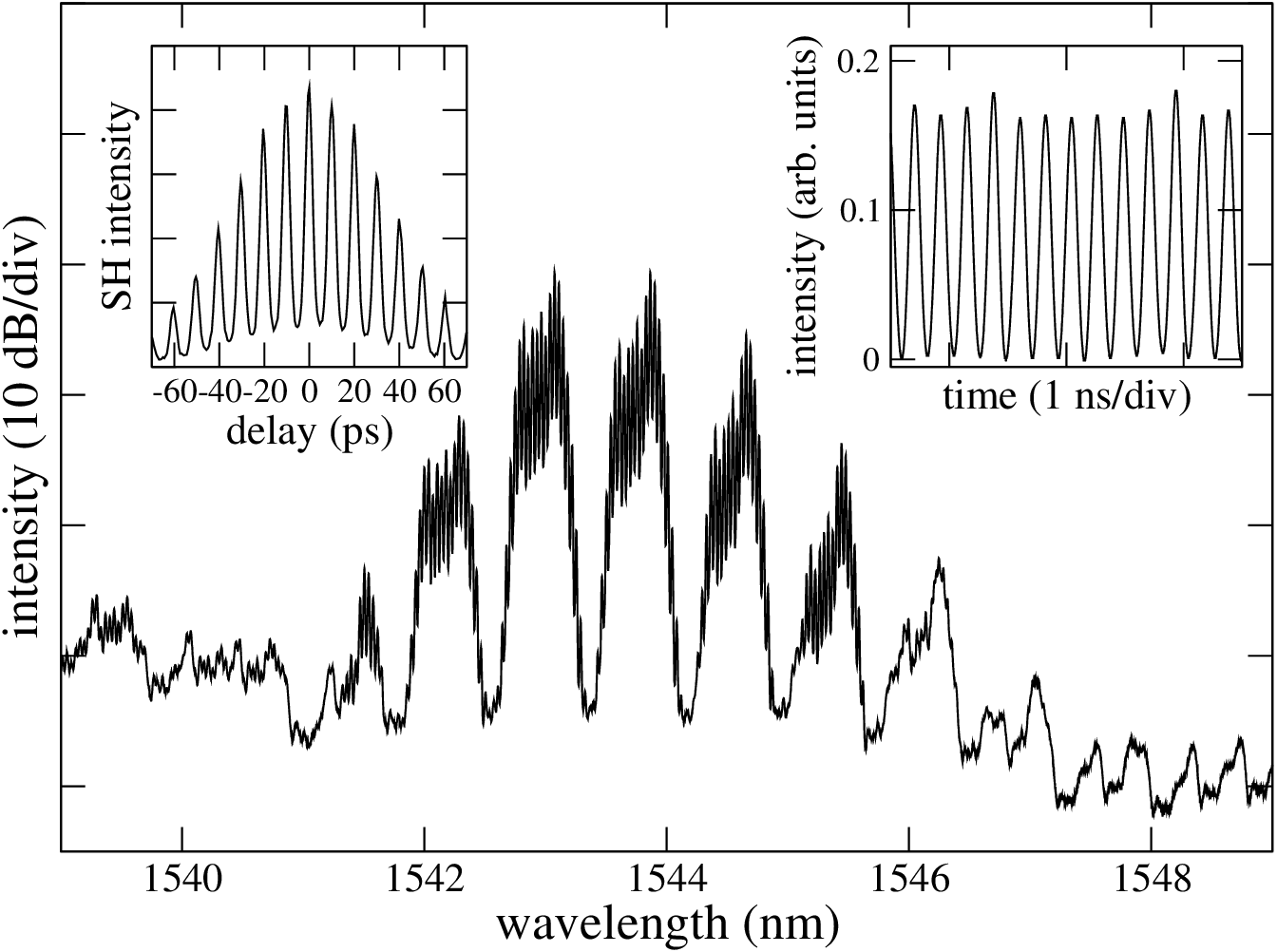}}
\caption{\label{expt_fig1} Optical spectrum for a  voltage of -2.15 V across the 
saturable absorber section.  The current in the gain section is 120 mA. 
Left inset: Autocorrelation measurement showing Q-switched mode-locking. Right inset:
Intensity time trace showing a large amplitude modulation of the total intensity.}
\end{figure}

Fig. \ref{expt_fig1} displays the optical spectrum obtained for a reverse bias 
of -2.15 V applied to the saturable absorber section of the device. In this case the 
dynamics are dominated by a set of strongly broadened modes that are spaced at 
twice the fundamental Fabry-Perot mode spacing. Interestingly,
we do not observe precisely six of these dominant modes but rather there are
five dominant modes with a well developed internal structure and a pair of much weaker
satellites. The autocorrelation measurement shown in the left inset indicates that the 
dynamical state of the device is Q-switched mode-locking in this case, where bunches of 
mode-locked pulses are emitted \cite{vasilev}. A measurement of the time trace of the total intensity 
of the laser is shown in the right inset, which shows a strong intensity modulation on 
a nanosecond time scale. This time scale determines the spacing of the individual peaks 
that make up each of the broadened Fabry-Perot modes.  

\begin{figure}[t]
\centerline{\includegraphics[width=8.0cm]{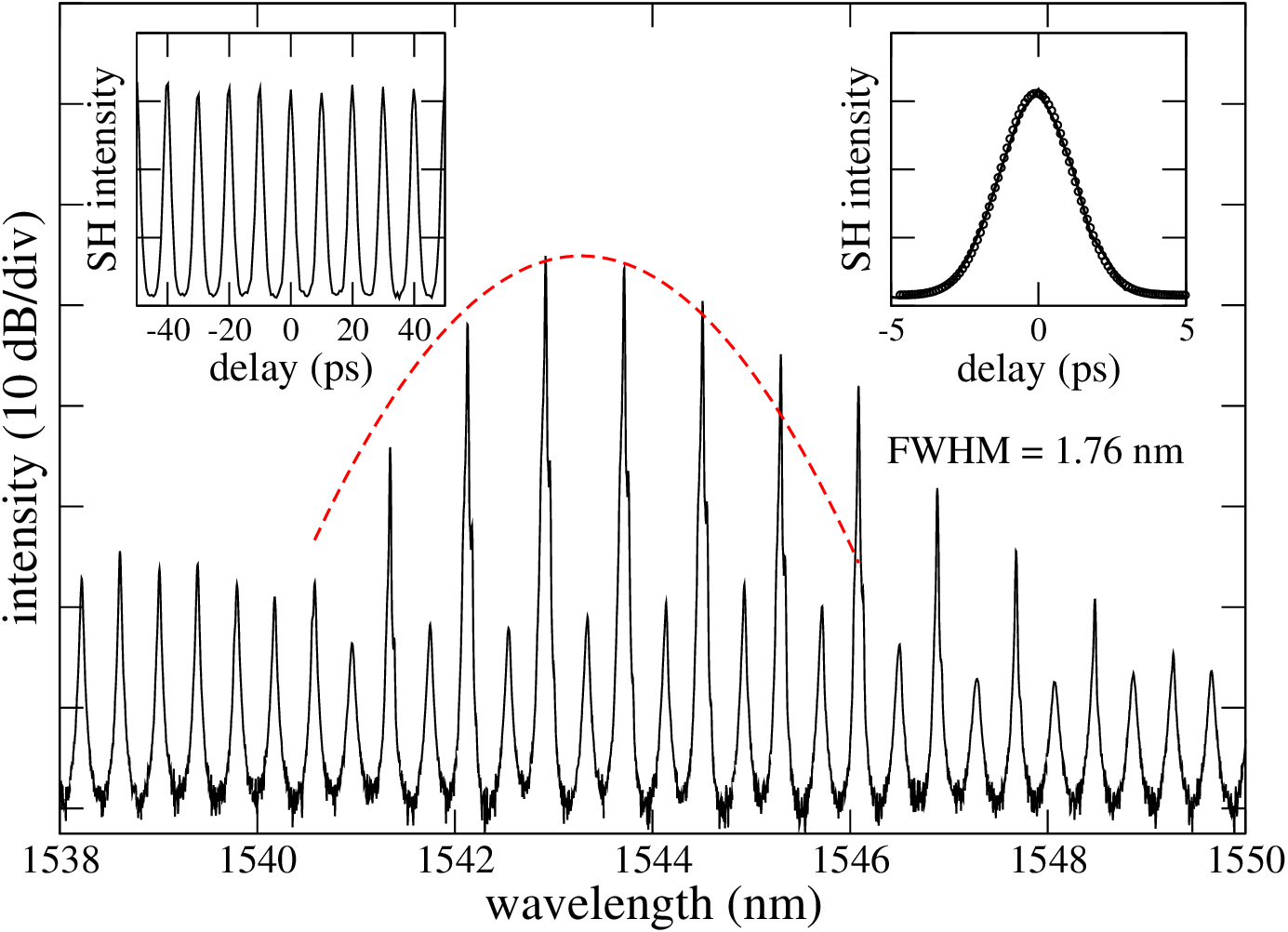}}
\caption{\label{expt_fig2} Optical spectrum for a  voltage of -2.20 V across the 
saturable absorber section. The current in the gain section is 120 mA. The dashed line
is a Gaussian fit to the spectrum of modes at the first harmonic over the bandwidth
indicated. Left inset: Autocorrelation measurement showing pulsed output at 100 GHz
repetition rate. Right inset: Autocorrelation measurement with Gaussian fit.}
\end{figure}

Mode-locking was observed with a voltage of -2.20 V applied to the saturable absorber 
section. The mode-locked spectrum is shown in Fig. \ref{expt_fig2} where 
a comb of  modes at a frequency separation of 100 GHz dominate 
the spectrum. In this case, more than the original six primary modes that 
were selected have significantly more power than the background modes. 
We attribute this broadening of the spectrum to power transfer by four-wave mixing.
The Gaussian fit to the intensity profile over a bandwidth of eight modes that was
used to calculate the full width half-maximum (FWHM) is shown in the figure. 

The  corresponding intensity autocorrelation is shown in the insets of Fig. 
\ref{expt_fig2}. The left inset demonstrates that well separated pulses are 
obtained at 100 GHz repetition rate, while on the right a Gaussian fit to a single 
pulse is shown. From the FWHM of the optical spectrum 
(1.76 nm) and the deconvolved pulsewidth of 2.06 ps, we obtain a time-bandwidth 
product $\Delta \nu \Delta \tau \sim 0.46$, which is close to the Fourier limited
value of $\sim 0.44$. We note that at 220 GHz, the FWHM of the optical spectrum 
is approximately equal to half of the bandwidth spanned by the comb of modes selected. 
Because the mode selection mechanism acts to predetermine the carrier wave frequency, 
we expect that shorter pulses may be obtained where the position of the central 
mode $m_{0}$ is optimized with respect to the material gain and loss dispersion. This 
conclusion is supported by the observation that otherwise identical Fabry-Perot devices
will mode-lock for different values of the drive parameters and with significantly
detuned carrier wave frequencies. 

We note also finally that while we have chosen to select a comb of six modes with 
uniform thresholds  in the device considered here, our approach gives great 
freedom to tailor the spectrum of the mode-locked device. The choice of spectral 
filtering function can be regarded as an additional degree of freedom that 
determines not only the repetition rate but also the pulse shape and duration. 
For example, the generation of pulses with approximately square profile should 
be possible, while some pulse shortening should be also possible through tailoring 
of the linear modal losses in order to compensate for the material gain 
dispersion. In addition, our approach will allow a minimal set of locked modes 
to be selected, which will enable optical synthesis of THz and millimeter wave 
frequencies to be achieved.  

In conclusion, we have demonstrated passive harmonic mode-locking of a comb of 
discrete Fabry-Perot modes in a diode laser that incorporates a non-periodic 
effective index profile. Near transform limited pulses of 2 ps duration were 
obtained at 100 GHz repetition rate. Our results demonstrate an interesting 
alternative approach to harmonic mode-locking in diode lasers with significant 
potential applications in optical waveform and frequency synthesis. 

\textit{Acknowledgments.} This work was supported by Science Foundation Ireland 
and Enterprise Ireland. The authors thank Eblana Photonics for the preparation 
of sample devices.

\end{document}